\documentclass[pra,letterpaper,aps,10pt,superscriptaddress,twocolumn,floatfix,showpacs,nobibnotes]{revtex4-2}

\usepackage{graphicx}
\usepackage{amsmath, amsfonts, amssymb}
\usepackage{bm}
\usepackage{float}
\usepackage{microtype}
\usepackage{bbm}
\usepackage{dsfont}
\usepackage{enumitem}
\usepackage{color, xcolor}
\usepackage{url}
\usepackage{braket}
\usepackage{upgreek}
\usepackage{subfigure}
\usepackage{comment}
\usepackage[most]{tcolorbox}  
\usepackage{xcolor}           
\usepackage{amsmath,amssymb}  

\usepackage{epsfig}
\usepackage{epstopdf}
\DeclareGraphicsExtensions{.pdf,.eps,.png,.jpg,.mps,.svg}

\usepackage[colorlinks, linkcolor=blue, citecolor=blue, urlcolor=blue, breaklinks=true]{hyperref}

\newcommand{\fref}[1]{Fig.~\ref{#1}}

\renewcommand{\eqref}[1]{Eq.~\textup{(\ref{#1})}}
\begin{document}

\title{Microscopic theory of a radiation-balanced solar laser}
    
\author{A. Jaber}
\affiliation{Department of Physics, University of Ottawa, Ottawa, ON K1N 6N5, Canada}
\affiliation{Max Planck Centre for Extreme and Quantum Photonics, Ottawa, ON K1N 6N5, Canada}

\author{M. Küblböck}
\affiliation{Max Planck Centre for Extreme and Quantum Photonics, Ottawa, ON K1N 6N5, Canada}
\affiliation{Max Planck Institute for the Science of Light, D-91058 Erlangen, Germany}
\affiliation{Department of Physics, Friedrich-Alexander-Universität Erlangen-Nürnberg, D-91058 Erlangen, Germany}

\author{J.-M. M\'{e}nard}
\affiliation{Department of Physics, University of Ottawa, Ottawa, ON K1N 6N5, Canada}
\affiliation{Max Planck Centre for Extreme and Quantum Photonics, Ottawa, ON K1N 6N5, Canada}
\affiliation{School of Electrical Engineering and Computer Science, University of Ottawa, Ottawa, ON K1N 6N5, Canada}

\author{H. Fattahi}
\affiliation{Max Planck Centre for Extreme and Quantum Photonics, Ottawa, ON K1N 6N5, Canada}
\affiliation{Max Planck Institute for the Science of Light, D-91058 Erlangen, Germany}
\affiliation{Department of Physics, Friedrich-Alexander-Universität Erlangen-Nürnberg, D-91058 Erlangen, Germany}

\author{C. Genes}
\affiliation{TU Darmstadt, Institute for Applied Physics, Hochschulstraße 4A, D-64289 Darmstadt, Germany}
\affiliation{Max Planck Centre for Extreme and Quantum Photonics, Ottawa, ON K1N 6N5, Canada}
\affiliation{Max Planck Institute for the Science of Light, D-91058 Erlangen, Germany}

\date{\today}

\begin{abstract}
We develop a microscopic open-quantum-system theory for a radiation-balanced solar laser (RBSL) based on ytterbium-doped yttrium aluminum garnet (Yb:YAG), in which optical gain, thermal redistribution among sublevels of the electronic ground and excited manifolds, and lattice-temperature dynamics are treated within a unified framework. Starting from a Lindblad master equation for a multilevel gain medium coupled to a cavity mode, we include incoherent solar pumping, spontaneous emission, cavity loss, and phonon-assisted intra-manifold relaxation obeying detailed balance. In the regime of fast thermalization within each electronic manifold, a compact temperature-dependent two-level model is derived, in which the gain, inversion, and lasing threshold are controlled by Boltzmann occupation factors and partition functions of the electronic sublevels. This microscopic reduction is then coupled self-consistently to a thermal balance equation accounting for anti-Stokes fluorescence cooling, quantum-defect heating, parasitic absorption, and heat exchange with the environment. The theory predicts several operating regimes, including pure cooling, lasing with net cooling, and lasing with net heating, as well as dynamical effects such as delayed lasing onset induced by self-cooling into threshold. In contrast to earlier radiation-balanced laser (RBL) models based mainly on macroscopic rate equations and thermodynamic balance arguments, the present approach provides a microscopic description of the feedback between quantum optical dynamics and temperature redistribution. It therefore offers a physically transparent framework for analyzing RBSLs and for identifying design strategies that exploit level structure, thermalization, and photonic-environment engineering to stabilize laser operation while minimizing internal heat load.

\end{abstract}

\maketitle

\section{Introduction}

Over the past decades, significant progress has been made in two seemingly distinct yet increasingly convergent fields: i) solar-pumped solid-state lasers and ii) anti-Stokes optical cooling of solids. Recently, there has been a proposal to combine these two aspects into a radiation-balanced solar laser (RBSL)~\cite{Kueblboeck2026Solar}.

Since the first demonstrations of sunlight-driven lasing using rare-earth-doped crystals~\cite{Lando1974SolarNdYAG, Arashi1984Solar18WNdYAG}, efforts have expanded toward efficient cavity geometries~\cite{Weksler1988SolarPumped,Lipson1996DesignOptimization,Lando2003SolarNdYAGHighCollection} and gain media optimized for direct solar pumping~\cite{Garcia2022CeNdYAG45Percent,Liang2025HighEfficiencySolarPumpedLasers} with a current record of $97~\mathrm{W}$ output power~\cite{Wang2026SolarCeNdYAG97W}. Recent analyses have revisited the practicality of solar lasers in space-based platforms, emphasizing optical filtering, radiative balance, and the use of broadband sunlight as an effective pump~\cite{Kueblboeck2024SolarLasers}. 

Anti-Stokes fluorescence cooling has emerged as a promising mechanism for passive, vibration-free solid-state refrigeration. Following its first experimental demonstration in ZBLAN glass~\cite{Epstein1995LaserCoolingZBLAN}, cooling has been extended to cryogenic regimes in Yb-doped fluoride crystals~\cite{Fernandez2000YbDopedGlassesCooling,Thiede2005OpticalRefrigeration208K,Patterson2008TmBaY2F8Cooling,Seletskiy2010YbYLF155K,Fernandez2012LowPhononMaterialsCooling,Volpi2018YbKYF4Cooling} and other crystals such as KPb$_2$Cl$_5$~\cite{Mendioroz2002YbKPb2Cl5Cooling,Fernandez2006BulkErbiumCooling,Fernandez2012LowPhononMaterialsCooling}, YAG~\cite{Nemova2014YbYAG}, CaF$_2$, SrF$_2$~\cite{Puschel2022YbCaFSrF2}, and even silica~\cite{Mobini2020SilicaGlassCooling,Mobini2021YbSilicaCooling}. Theoretical proposals continue to push the fundamental temperature limits of this process~\cite{ToledoTude2024OvercomeLimits}.\\
\begin{figure}[t]
\includegraphics[width=.9\columnwidth]{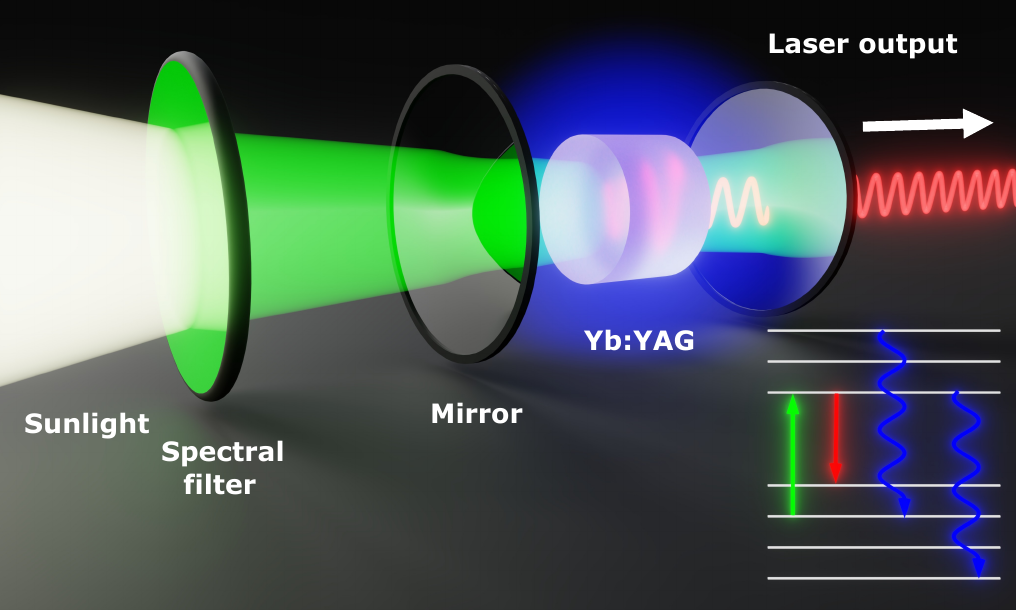}
\caption{\textit{Depiction of a RBSL using ytterbium-doped yttrium aluminum garnet (Yb:YAG) as the gain medium.}
Spectrally filtered sunlight enters the laser cavity through the left mirror to provide a narrow-band incoherent pump for a Yb:YAG crystal placed inside the optical mode volume. The filtered solar pump power (green) is converted into laser output power (red), fluorescence power (blue), and residual thermal power. The pumping scheme can be engineered to enhance anti-Stokes fluorescence from the excited manifold, thereby combining solid-state optical cooling with laser operation.}
\label{Fig1}
\end{figure}
A natural next step is to integrate these technologies: using solar light not only as an energy source for lasing, but also to enable the stabilization of gain medium temperature via radiative cooling (a mechanism involving radiative redistribution of energy~\cite{Vogl2009LaserCoolingByCollisionalRedistribution,Sommer2019LaserRefrigerationHollowCore}). This approach forms the core idea of RBSLs---devices capable of maintaining thermal equilibrium without external cooling infrastructure. Partial progress has been made in analyzing the possibility of achieving a thermal laser, albeit not fed by sunlight. 
The radiation-balanced concept was first formulated for a single-pass optical amplifier~\cite{Bowman1999NoInternalHeat}, rather than a laser oscillator. It was subsequently extended to cavity-based systems through studies of lasing stability and continuous-wave radiation-balanced laser configurations~\cite{Bowman2002RBLStability,Li2004RBLCWOscillators}. The first attempts and demonstrations of radiation-balanced laser (RBL) operation utilized Yb:KGW~\cite{Bowman2005ReducedThermalLoading} and Yb:YAG~\cite{Bowman2010MinimizingHeat}. Recent experimental work typically focuses on Yb:YAG disks as gain media~\cite{Yang2019RBYbYAGDisk,Wu2023AmbientTempYbYAG} and Yb-doped silica fiber lasers~\cite{Knall2021RBSilicaFiberLaser,Balliu2024RBSilicaFiber200mW}.
Existing theoretical studies of solar lasers, RBLs, and RBSLs have so far relied on macroscopic semiclassical laser rate equations to describe their operating principles. A microscopic quantum-optical theory of these devices remains undeveloped.

We explore a direct solar irradiance-to-laser light device capable of operating across a broad and realistic temperature range while counteracting inadvertent heating produced during operation.
This device filters solar light into a narrow-band incoherent pump incident on a laser gain medium inside an optical cavity. To maintain stable operation, such a system must be temperature controlled, which necessitates consideration of passive cooling mechanisms such as the redistribution of produced heat into spontaneous emission which is radiated away. A schematic of the basic operation is shown in Fig.~\ref{Fig1}. Filtered sunlight is incident on a gain medium placed within a laser cavity. The spectral filtering is accomplished either with the explicit use of a narrow-band filter around the pump wavelength of interest or due to the absorption spectrum of the gain medium peaking at suitable pump wavelengths and being negligible otherwise, which is common with solid-state laser gain media \cite{Liang2025HighEfficiencySolarPumpedLasers}. Apart from the laser output produced by the pump, the gain medium will fluoresce due to spontaneous emission from non-lasing transitions. Engineering the Stokes and anti-Stokes transition landscape of the solid-state gain medium can lead to this fluorescence power contributing to a net cooling of the gain medium. 

As the gain medium, we consider Yb:YAG, consisting of a YAG crystal in which a small fraction of Y lattice sites is replaced by $\mathrm{Yb}^{3+}$ ions. YAG is a widely used host material because of its broad optical transparency, high thermal stability, and absence of birefringence. Since the doping we consider is sufficiently dilute (typically $1$--$5\%$), inter-ion interactions can be neglected, and the relevant microscopic description reduces to the level structure of a single $\mathrm{Yb}^{3+}$ ion embedded in the host crystal. At the same time, the thermal properties of YAG remain essential, since they determine both inadvertent heating through residual absorption and cooling through anti-Stokes fluorescence, where the energy of absorbed pump photons is redistributed into spontaneously emitted photons of higher energy. Our aim is to identify operating regimes that maximize laser output while maintaining the gain-medium temperature within a range that is not detrimental to the host matrix.

Our approach systematically reduces a full open quantum system description to a minimal, analytically tractable model while retaining the essential physics of coupled optical and thermal dynamics. We begin with a multi-level Lindblad framework, which we numerically simulate in order to benchmark all subsequent approximations. The central step is the adiabatic elimination of fast intra-manifold thermalization, justified when the redistribution rate $\Gamma$ dominates over spontaneous emission $\gamma$, cavity decay $\kappa$, and pump rate $R$. This reduction yields a temperature-dependent two-manifold Maxwell--Bloch model, in which gain and inversion depend explicitly on thermal population factors through the corresponding partition functions. Building on this reduced description, we derive a closed set of coupled equations for the cavity field, medium polarization, and total excitation, and embed them into a dynamical thermal balance equation that consistently includes anti-Stokes cooling, quantum-defect heating, parasitic absorption, and environmental coupling. The resulting minimal model links optical dynamics and temperature evolution, enabling analytical expressions for the lasing threshold, output power, and steady state temperature, while also making clear the regime of validity of the reduction.

Our calculations show that lasing is only achievable when the spontaneous-emission rate $\gamma$ is sufficiently small, so that the lasing threshold can be reached at realistic pump powers. In practice, this strongly favors rare-earth ions, whose parity-forbidden optical transitions naturally provide long excited-state lifetimes and correspondingly small $\gamma$. For emitters with larger spontaneous-emission rates, the pump powers required to overcome threshold become prohibitively high, making RBL operation essentially unattainable under realistic conditions.

Within this framework, we predict three operating regimes for the same physical device: (i) net cooling without lasing, (ii) radiation-balanced lasing with simultaneous cooling, and (iii) lasing with net heating. In contrast to standard RBL scenarios, the transitions between these regimes are governed not only by global heat balance but also by temperature-dependent redistribution among internal sublevels. This leads to a temperature-dependent lasing threshold and gives rise to distinctive signatures such as delayed lasing onset through self-cooling and an anticorrelation between broadband fluorescence and coherent laser output near thermal equilibrium.

\section{The model}

We begin by describing the physical platform under consideration. We then introduce the formalism for open system dynamics, which includes Hamiltonian dynamics for the dopant material interacting with external light sources and the optical resonator. Finally, we incorporate loss processes such as radiative processes, optical pumping, and thermalization owing to the contact of the dopant with the solid-state matrix.

\subsection{Dopant and host medium}

Laser transitions of $\mathrm{Yb}^{3+}$ occur within the 4f orbital, which has orbital angular momentum \(L = 3\). With electron spin \(S = 1/2\), spin-orbit coupling produces total angular momentum states \(J = 5/2\) and \(J = 7/2\), which differ by \(\Delta J = 1\) and thus allow dipole transitions. According to Hund’s rules, the level with the larger \(J\) lies lower in energy. Therefore, the \(J = 7/2\) states form the ground-state manifold, and the \(J = 5/2\) states form the excited-state manifold.

Due to the crystal field (similar to a static Stark shift) generated by the YAG environment, the degeneracy of each \(J\) level is partially lifted, creating manifolds of distinct sublevels. The total degeneracies are \(2J + 1 = 8\) for \(J = 7/2\) and \(6\) for \(J = 5/2\). Time-reversal symmetry (Kramers' theorem) ensures that degeneracy is lifted only partially, leaving 4 and 3 Kramers' doublets in the ground- and excited-state manifolds, respectively~\cite{Sakurai_Napolitano_2020,auzel2002relationship,demirkhanyan2022evidence,Barry2024Elucidating}. Thus, a $\mathrm{Yb}^{3+}$ ion can be represented with a 7-level model. 

\begin{figure}[t]
\includegraphics[width=0.9\columnwidth]{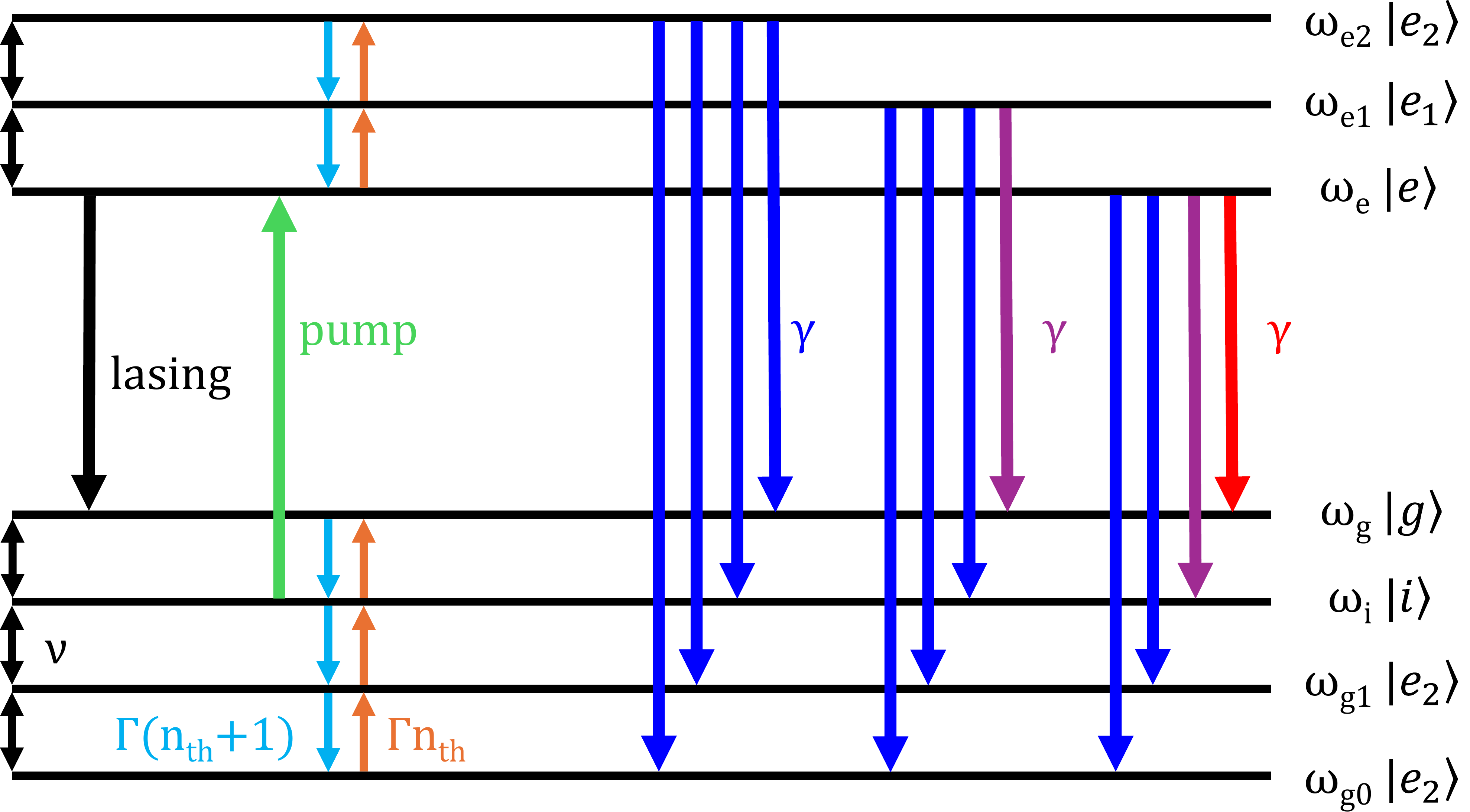}
\caption{
 \textit{Energy level diagram of a single $\mathrm{Yb}^{3+}$ ion.} The electronic levels are grouped into ground and excited manifolds, separated by a large optical transition frequency (hundreds of THz). Smaller intra-manifold splittings of about 5~THz enable fast thermalization, effectively maintaining a Boltzmann distribution at all times. The simplified model assumes equal energy spacing, uniform thermalization rates \(\Gamma\), and identical transition dipole moments, resulting in a common spontaneous emission rate \(\gamma\). Detuning between the pump frequency and the manifold transitions enables net energy extraction from phonons via anti-Stokes emission. The spontaneous transitions are color-coded to indicate whether the transition is Stokes (red), anti-Stokes (blue), or neutral (purple).
}
\label{Fig2}
\end{figure}

The separation between the two manifolds is approximately \(322.5~\mathrm{THz}\), while the intra-manifold splittings are around \(5~\mathrm{THz}\). As a first step, we model the complete level scheme of a single $\mathrm{Yb}^{3+}$ ion in an incoherently pumped cavity, treating it as an open quantum system. The energy level diagram, including all relevant transitions, is shown in Fig.~\ref{Fig2}.

The ground-state manifold includes four levels with transition frequencies \(\omega_{g_0}, \omega_{g_1}, \omega_{g_2} = \omega_{i}, \omega_{g_3} = \omega_g\), while the excited-state manifold includes three levels with \(\omega_{e_0} = \omega_{e}, \omega_{e_1}, \omega_{e_2}\). Transitions between the electronic manifolds are at optical frequencies and stem from:
\begin{itemize}
    \item Spontaneous emission from any excited sublevel to any ground sublevel
    \item External pumping via incoherent sunlight on the transition $\ket{i} \rightarrow \ket{e}$
    \item Coupling to the cavity mode on the lasing transition $\ket{g} \rightarrow \ket{e}$.
\end{itemize}

Within each manifold, we include thermalization processes with rates \(\Gamma \overline{n}\) (heating) and \(\Gamma (\overline{n}+1)\) (cooling), where the mean phonon occupation number \(\overline{n}\) follows the Bose–Einstein distribution:
\begin{equation*}
    \overline{n} = \frac{1}{e^{\hbar \omega / k_B T} - 1},
\end{equation*}
with \(\hbar\) being the reduced Planck constant, \(\omega\) being the frequency difference between two adjacent sublevels, \(k_B\) being the Boltzmann constant, and \(T\) being the temperature in Kelvin. This ensures that the system, when pushed out of equilibrium, thermalizes at rate \(\Gamma\) to the externally imposed temperature \(T\).

\subsection{Hamiltonian terms}

The system includes the internal level structure of the ions, the quantized cavity field, and their mutual interaction. The total Hamiltonian is given by
\begin{equation}
    \hat{\mathcal{H}}(t) = \hat{\mathcal{H}}_{0} + \hat{\mathcal{H}}_c + \hat{\mathcal{H}}_{\text{int}},
    \label{eq:H_TD}
\end{equation}
where each term represents a physical contribution:

\begin{itemize}
    \item \emph{electronic free Hamiltonian}
    \begin{equation}
        \hat{\mathcal{H}}_{0} = \sum_{k=0}^{3} \hbar \omega^{(g)}_k \ket{g_k}\bra{g_k} + \sum_{k=0}^{2} \hbar \omega^{(e)}_k \ket{e_k}\bra{e_k},
    \end{equation}
    with \(\hbar\omega^{(g)}_k\) and \(\hbar\omega^{(e)}_k\) denoting the energy of all sublevels in the two electronic manifolds.

    \item \emph{cavity Hamiltonian}
    \begin{equation}
        \hat{\mathcal{H}}_c = \hbar \omega_c \hat{a}^\dagger \hat{a},
    \end{equation}
    where \(\omega_c = \omega_e - \omega_g\), and \(\hat{a}^\dagger, \hat{a}\) are the photon creation and annihilation operators.

    \item \emph{light–matter interaction}
    \begin{equation}
        \hat{\mathcal{H}}_{\text{int}} = \sum_{j=1}^N \hbar g_j \left( \hat{a}^\dagger \hat{\sigma}^{(j)} + \hat{a} \hat{\sigma}^{\dagger(j)} \right),
    \end{equation}
    where \(g\) is the coupling strength and \(\hat{\sigma} = \ket{g}\bra{e}\) is the lowering operator for the lasing transition.

\end{itemize}
\subsection{Incorporating loss mechanisms: the master equation approach}

To compute the time evolution of level populations, we must account for open system effects and the possible mixed-state nature of the system. Using the density matrix formalism in the Schrödinger picture, the dynamics are governed by a master equation of Lindblad form:
\begin{equation}
    \frac{d\hat{\rho}}{dt} = -\frac{i}{\hbar } [\hat{\mathcal{H}}, \hat{\rho}] 
    + \sum_i \mu_i \left( \hat{J}_i \hat{\rho} \hat{J}_i^\dagger 
    - \frac{1}{2} \left\{ \hat{J}_i^\dagger \hat{J}_i, \hat{\rho} \right\} \right),
    \label{eq:master_rho}
\end{equation}
where the first term describes coherent evolution and the second captures dissipative processes. The operators \(\hat{J}_i\) are quantum jump (collapse) operators with decay rates \(\mu_i\).

The relevant jump processes are:
\begin{itemize}
    \item \textit{Solar pump} acting between the ground and electronic manifolds with collapse operator \(\hat{\sigma}^\dagger_p = \ket{e}\bra{i}\), and pump rate $R$.
    \item \textit{Spontaneous emission} acting between the electronic manifolds with collapse operator \(\hat{J}_i = \ket{g_k}\bra{e_j}\), and emission rates \(\gamma_{jk} = \gamma\).
    \item \textit{Thermalization:} acting within each manifold to push towards thermal equilibrium among all sublevels with collapse operator \(\hat{J}_i = \ket{g_k}\bra{g_j}\) or \(\ket{e_k}\bra{e_j}\), and with rates \(\Gamma \bar{n}\) and \(\Gamma (\bar{n} + 1)\).
    \item \textit{Cavity photon loss:} loss via cavity mirrors with collapse operator \(\hat{J}_i = \hat{a}\), and photon loss rate \(\kappa\).
\end{itemize}

Expectation values for any generic operator $\hat O$ evolve as
\begin{equation*}
    \frac{d}{dt} \langle \hat{O} \rangle = \mathrm{Tr} \left[ \hat{O} \, \frac{d\hat{\rho}}{dt} \right].
\end{equation*}
We will use this equation coupled with assumptions regarding factorization of two operator expectation values (mean-field approximation) in order to obtain a set of equations of motion for all variables of interest (field amplitude, populations, and coherences).

\section{Equations of motion}

We derive equations of motion for the full seven-level system, which serves as a reference model against which reduced descriptions are benchmarked. The gain medium is treated as an ensemble of \(N\) identical ions interacting with a single cavity mode. All ions are assumed to contribute independently and identically to the macroscopic polarization. We introduce the following notations:
\begin{itemize}
  \item The cavity field amplitude is
  \[
    \alpha = \langle \hat{a} \rangle .
  \]
  \item The optical coherence on the lasing transition for a single ion is
  \[
    s = \langle \hat{\sigma}_{ge} \rangle .
  \]
  \item The populations of the ground manifold levels are
  \(
    p_{g_0},\, p_{g_1},\, p_i,\, p_g
  \),
  and those of the excited manifold are
  \(
    p_e,\, p_{e_1},\, p_{e_2}.
  \)
\end{itemize}

After summing over all ions, we define the macroscopic quantities
\begin{itemize}
  \item Macroscopic polarization:
  \[
    S = \sum_{j=1}^{N} s^{(j)} .
  \]
  \item Macroscopic populations:
  \[
    P_\ell = \sum_{j=1}^{N} p_\ell^{(j)}, \qquad \ell \in \{ g_0,g_1,i,g,e,e_1,e_2 \}.
  \]
\end{itemize}

\subsection{Ensemble of \(N\) identical ions}

In the ensemble model of Yb:YAG, the macroscopic populations represent the total number of ions in each respective energy level. The cavity field evolves according to
\begin{equation}
\dot{\alpha}
= -\frac{\kappa}{2}\,\alpha + g\,S ,
\end{equation}
where \(\kappa\) is the cavity decay rate and \(g\) is the single-ion coupling strength. The gain term arises from the macroscopic polarization of the ensemble.

The macroscopic polarization on the lasing transition obeys
\begin{equation}
\dot{S}
= g\,\alpha\,(P_e - P_g)
- \left[ \tfrac{1}{2}\Gamma(2\bar n + 1) + 2\gamma \right] S ,
\end{equation}
where \(\bar n\) is the mean phonon occupation number and the damping terms account for phonon-induced dephasing and spontaneous emission.

The populations of the ground-manifold levels evolve as
\begin{subequations}
\begin{align}
\dot{P}_{g_0} =&
\gamma (P_e + P_{e_1} + P_{e_2})
- \Gamma \bar n\, P_{g_0}
+ \Gamma (\bar n + 1)\, P_{g_1}, \\[4pt]
\dot{P}_{g_1} =&
\gamma (P_e + P_{e_1} + P_{e_2})\notag \\
& + \Gamma \bar n\, (P_{g_0}-P_{g_1}) 
 + \Gamma (\bar n + 1)\, (P_i-P_{g_1}) , \\[4pt]
\dot{P}_i =&
\gamma (P_e + P_{e_1} + P_{e_2})
- R\,P_i  \notag \\
& + \Gamma \bar n\, (P_{g_1}-P_i)
+ \Gamma (\bar n + 1)\, (P_g-P_i) , \\[4pt]
\dot{P}_g =&
\gamma (P_e + P_{e_1} + P_{e_2})
+ g(\alpha^\ast S + \alpha S^\ast) \notag \\
& + \Gamma \bar n\, P_i 
 - \Gamma (\bar n + 1)\, P_g .
\end{align}
\end{subequations}
Here \(R\) denotes the incoherent pump rate from the intermediate level \(i\) into the excited level \(e\).

The excited-manifold populations obey
\begin{subequations}
\begin{align}
\dot{P}_e =&
R\,P_i
- g(\alpha^\ast S + \alpha S^\ast)
- 4\gamma\, P_e  \notag \\
& - \Gamma \bar n\, P_e
+ \Gamma (\bar n + 1)\, P_{e_1} \\[4pt]
\dot{P}_{e_1} =&
- 4\gamma\, P_{e_1} \notag\\
& + \Gamma \bar n\, (P_e-P_{e_1})
 + \Gamma (\bar n + 1)\, (P_{e_2}-P_{e_1})
 , \\[4pt]
\dot{P}_{e_2} =&
- 4\gamma\, P_{e_2} \notag\\
&+\Gamma \bar n\, P_{e_1} - \Gamma (\bar n + 1)\, P_{e_2} .
\end{align}
\end{subequations}
The total population
\begin{equation}
P_{g_0} + P_{g_1} + P_i + P_g + P_e + P_{e_1} + P_{e_2} = N
\end{equation}
is conserved by the dynamics. This full seven-level model constitutes the reference description used to validate the reduced model introduced in a following section.

\subsection{Temperature as a dynamical parameter}

We introduce a phenomenological rate equation to model the interplay between pump-induced heating and spontaneous emission cooling:
\begin{equation}
    m_{\text{host}} C_{\text{host}} \frac{dT}{dt} = \mathcal{P}_{\text{heat}} - \mathcal{P}_{\text{cool}}- \mathcal{P}_{\text{env}},
\end{equation}
where $m_{\text{host}}$ and $C_{\text{host}}$ are the mass and heat capacity of the host, $\mathcal{P}_{\text{heat}}$ accounts for the phonon energy deposited via quantum defect and Beer-Lambert heating,  $\mathcal{P}_{\text{cool}}$ accounts for phonon energy extracted via anti-Stokes fluorescence, and $\mathcal{P}_{\text{env}}$ accounts for the energy exchange of YAG with the environment through conduction, convection, and radiation. It is convenient to rewrite the temperature dynamics in terms of $\bar n(t)$:
\begin{align}
\frac{d\bar n}{dt}= \chi(\mathcal{P}_{\text{heat}}-\mathcal{P}_{\text{cool}}-\mathcal{P}_{\text{env}})\bar n \bigl(\bar n+1\bigr)
     \left[\ln\!\left(1+\frac{1}{\bar n}\right)\right]^{\!2}, \label{eq:Temp_dynamics}
\end{align}
where we have defined
\[
\chi \equiv \frac{k_B}{m_{\text{host}}C_{\text{host}}\hbar\nu}.
\]
We will now detail the processes leading to either heating or cooling of the material and the environment.
\subsubsection{Cooling processes} 
Cooling occurs with the up-conversion of energy from the pump frequency into higher spontaneous-emission photons (see Fig.~\ref{Fig2}). Our choice in selecting the pump transition (wavelength) is to ensure there are more anti-Stokes pathways than Stokes pathways. For every spontaneous emission $\ket{e_j}\!\to\!\ket{g_k}$  
\[
  \Delta E_{jk}= \hbar(\omega_{e_j}-\omega_{g_k}-\omega_p)
               = \hbar\,(j+k-1)\,\nu,
\]
with $j\in\{0,1,2\},\;k\in\{0,1,2,3\}$. Assuming an identical partial rate $\gamma$ on every optical branch,
the cooling power carried by that branch equals rate $\times$ energy, written as
\[
  \mathcal{P}_\text{cool}^{(jk)}= \gamma\,P_{e_j}\,\Delta E_{jk},
\]
where \(\Delta E_{jk}\) is the transition energy of each spontaneous emission. We can calculate the phonon energy taken from the lattice for each spontaneous transition. For a fixed excited sublevel $j$, we sum over the available ground sublevels $k$, and account for the pump transition adding one phonon of energy to the lattice:
\[
  \sum_{k=0}^{3} (j+k-1)
  = \begin{cases}
      2, & j=0 \\[2pt]
      6, & j=1 \\[2pt]
      10,& j=2
    \end{cases}
\]
The total cooling power is obtained by summing over all spontaneous transitions from the three excited states such that
\[
\mathcal{P}_{\mathrm{cool}}
  = \gamma\hbar \nu \Phi(\bar n)P_e
\]
where the temperature-dependent factor is given by
\[
 \Phi(\bar n)=2 \frac{(1 + 5\bar{n} + 9\bar{n}^2)}{(\bar{n}+1)^2}.
\]
The cooling power strongly depends explicitly on temperature through $\bar n$ and implicitly via $P_e$; the pump rate \(R\) also has an implicit contribution. We reiterate that $\mathcal{P}_{\mathrm{cool}}$ is the power exchanged with the lattice due to spontaneous emission. To calculate the total fluorescence power we add the contribution of photon energy leaving the system, giving $\mathcal{P}_{\mathrm{fluo}}=(4\gamma N_e)\hbar\omega_p+\mathcal{P}_{\mathrm{cool}}$.

\subsubsection{Heating processes} 

For the case of Yb:YAG in particular (typical to other ion-doped solid-state gain media), we can assume the four ground sublevels and three excited sublevels
rethermalize on a timescale much shorter than any optical or cavity
process.  Detailed balance between the phonon-assisted upward and downward transitions within each manifold, occurring at rates \(\Gamma \bar{n}\) and \(\Gamma (\bar{n}+1)\), respectively, guarantees that the net phonon flux into and out of the crystal vanishes at thermal equilibrium. The lattice neither heats nor cools through intra‑manifold phonon exchange once the
sublevel ladder has reached equilibrium.\\

The remaining heat sources are therefore optical: i) inherent heating owing to the down-conversion of pump photons into lasing photons, ii)  heating by absorption in YAG, and iii) heating by redistribution of pump photons into spontaneous emission at lower frequencies (taken into account with $\mathcal{P}_\text{cool}$).\\

The efficiency of the lasing process can be estimated from the coherent output power of the lasing cavity, $\mathcal{P}_\text{out}=\kappa \hbar \omega_c |\alpha|^2$. For each pump photon that successfully produces a lasing photon, a phonon of energy $\hbar\nu$ is deposited in the lattice as heat. We can then estimate this quantum defect power using the fraction $\nu/(\omega_c+\nu)\approx \nu/\omega_c$ so that
\[
  \mathcal{P}_\text{heat}^{(1)}
  = \kappa \hbar \nu  |\alpha|^2.
\]
 
The next term involves absorption in the material. This is proportional to the absorption coefficient $\zeta$ at the pump frequency and the material depth $L$. Given that we consider ion-doped gain media that are often thin disks, we will assume a small medium length $L$, allowing us to state the Beer-Lambert absorbed heating power as 
\[
  \mathcal{P}_\text{heat}^{(2)}
    = \zeta L \mathcal{P}_\text{in}.
\]

Finally we can add everything into $\mathcal{P}_\text{heat}$, which is then counteracted by anti‑Stokes fluorescence, so even without the presence of the environment, the system is temperature stabilized when \(\mathcal{P}_{\mathrm{heat}}=\mathcal{P}_{\mathrm{cool}}\). If this radiation-balanced condition is met when there is a non-zero laser power output, then the system will function as a RBSL.

\subsubsection{Environment contribution} 
In addition to the internal optical heating and cooling processes discussed above, the YAG host exchanges heat with its environment. In general this occurs through conduction to a mount or heat sink, convection to surrounding gas, and thermal radiation. For the thin-disk geometries considered here, conductive coupling to the mount dominates convection and radiation processes by orders of magnitude, allowing the environmental heat exchange to be approximated as $\mathcal{P}_{\text{env}}=G_\text{th}(T-T_\text{env})$, where $G_\text{th}$ is the thermal conductance. Order-of-magnitude estimates for the three mechanisms are provided in appendix~\ref{app:enviroheat}.
\section{RESULTS}

The full system of seven electronic levels coupled to the cavity field is kept as a means to benchmark results obtained in a simplified two-level model. However, because the reduced two‑level description provides clearer analytical access to the relevant field–matter interactions and associated performance regimes, our physical interpretation and threshold estimates rely on this simplified model, which we now introduce and justify. The simplified model emerges directly from the separation of timescales between phonon-assisted relaxation processes within each electronic manifold (thermalization) and all other processes. In particular, we assume
\[
\Gamma \gg \{\gamma,\kappa,R\},
\]
so that populations within each manifold relax practically instantaneously to thermal equilibrium. This assumption is typical in the RBL literature on ion-doped gain media \cite{Bowman1999NoInternalHeat,Bowman2010MinimizingHeat,Yang2019RBYbYAGDisk,Wu2023AmbientTempYbYAG,Li2004RBLCWOscillators,Bowman2002RBLStability}. More specifically, using the Boltzmann factor
\[
q \equiv \frac{\bar n}{1+\bar n},
\]
the excited-manifold populations satisfy
\[
P_{e_1} = q\,P_e, \qquad
P_{e_2} = q^2\,P_e,
\]
while the lower ground-manifold populations obey
\[
P_{g_1} = \frac{1}{q}\,P_i, \qquad
P_{g_0} = \frac{1}{q^2}\,P_i.
\]
It is then natural to introduce the total excited-manifold population
\[
N_e \equiv P_e + P_{e_1} + P_{e_2},
\]
which serves as the single population variable that will enter our equations, as the total ground-manifold population $N_g$ and excited manifold populations will add to $N_e+N_g=N$. Defining the excited- and ground-manifold partition functions
\[
Z_e = 1 + q + q^2, \qquad
Z_g = 1 + q + q^2 + q^3,
\]
the populations required for the optical dynamics are reconstructed algebraically as
\begin{equation}
P_e = \frac{N_e}{Z_e}, \qquad
P_g = \frac{q^3}{Z_g}\,(N - N_e), \qquad
P_i = \frac{q^2}{Z_g}\,(N - N_e).
\end{equation}
Because the pump is incoherent and all other coherences decay on the fast thermalization timescale, the only coherence retained is the coherence associated with the lasing transition \(S\). Substituting the above relations into the full equations of motion yields a closed set of equations for the reduced variables \(\{\alpha,S,N_e\}\) (derived in appendix~\ref{app:7to2}). To this, we add \eqref{eq:Temp_dynamics}, re-expressed in terms of the mean occupation number. This leads to our working set of equations, microscopically derived to characterize the time dynamics of the lasing medium and the optical cavity field amplitude:
\begin{widetext}
\begin{subequations}
\label{eq:strict_twostate}
\begin{align}
\dot{\alpha} &=
-\frac{\kappa}{2}\,\alpha
+ g\,S,
\\[6pt]
\dot{S} &=
g\,\alpha\,(P_e-P_g)
- \Gamma_S(\bar n)\,S,
\\[6pt]
\dot{N}_e &=
R\,P_i
- 4\gamma\,N_e
- g\!\left(\alpha^{*}S+\alpha S^{*}\right), \\[6pt]
\dot{\bar{n}} &= \chi\,\Bigg[\hbar \nu \Bigg(\kappa |\alpha|^2-\gamma \Phi(\bar n)\frac{N_e}{Z_e}\Bigg)+\zeta L\mathcal{P}_{\mathrm{in}}-G_\text{th}(T-T_\text{env})\Bigg]\,\bar n(\bar{n} + 1)
     \left[\ln\left(1 + \frac{1}{\bar{n}}\right)\right]^2.
\end{align}
\end{subequations}
\end{widetext}
This is a microscopic model for a lasing system in which the cavity field is driven by the collective coherence $S$, while $S$ itself is generated by the population inversion $N_e$. The description is at the mean-field level and allows one to identify a lasing threshold via a linear stability analysis around the trivial (non-lasing) solution, i.e., by introducing an infinitesimal seed. \textit{The threshold is reached when an arbitrarily small seed in the cavity field amplitude is exponentially amplified and evolves toward a finite steady state value that is independent of the initial seed magnitude.} Temperature enters implicitly through the system variables and explicitly through the effective damping rate
\[
\Gamma_S = \Gamma(\bar n + 1/2) + 2\gamma.
\]

\subsection{Threshold: analytical scalings}

At fixed temperature (hence fixed $\bar n$), the reduced two-manifold Maxwell--Bloch system admits an analytic threshold condition. In steady state with $\alpha\neq 0$, the field and polarization equations imply that the macroscopic coherence is proportional to the cavity field,
\begin{equation}
S=\frac{\kappa}{2g}\,\alpha,
\end{equation}
and consistency with the polarization equation yields a clamped population inversion on the lasing transition,
\begin{equation}
\Delta P_{eg}
\equiv P_e-P_g
=\frac{\kappa \Gamma_S}{2g^{2}}
.
\label{eq:DeltaPth_main}
\end{equation}

Threshold is reached when there is a non-zero field within the cavity at the lasing transition. One condition for lasing is the presence of photons in the cavity, requiring \(|\alpha|^2>0\). We can set the reduced equations of motion to steady state and solve for this condition (see appendix~\ref{app:thresholds}). This leads to a threshold value for the incoming pump rate
\begin{equation}
R_{\mathrm{th}}
=
\frac{4\gamma Z_e\!\left(\Delta P_{eg}\, Z_g + N q^3\right)}
{q^2\!\left(N - \Delta P_{eg}\, Z_e\right)}.
\label{eq:SI_Rth}
\end{equation}

This expression is only valid as long as $N>\Delta P_{eg}\, Z_e$, keeping $R$ positive and finite. This validity requirement comes from our constraint that there is a positive cavity field amplitude, $|\alpha|>0$, which is only true when above threshold, i.e. when the pump rate and number of gain ions is large enough. The analytical expression is an approximation that fits very well with the numerical simulation, above the threshold pump rate \(R_{\mathrm{th}}\). Also, notice that the threshold rate depends weakly on temperature via the variable $Z_e$, which varies from $1$ at low temperature to $3$ at very high temperature.

To consistently relate the incoherent pumping to the available solar
power, we impose a constraint at the level of the \emph{total excitation
flux} rather than treating the single-particle pump rate \(R\) as an
independent parameter. In the equations of motion, pumping injects
excitations into the gain medium at a rate \(R P_i\).

We determine this flux by equating the absorbed optical power to the
energy transferred into electronic excitations. This yields $\mathcal{P}_{\text{in}}=\hbar \omega_pR(t)\,P_i(t)$, which enforces that the total excitation rate is limited by the incident
photon flux. For convenience, this can be expressed in terms of an
effective population-dependent rate
\begin{equation}
    R(t)=\frac{\mathcal{P}_{\text{in}}}{\hbar\omega_p\,P_i(t)}.
\end{equation}

The physically meaningful quantity is the product
\(R P_i\), which remains finite and directly determined by the available
input power.

This formulation guarantees that the optical excitation dynamics remain
consistent with energy conservation and prevents unphysical regimes in
which the emitted laser power exceeds the incident power. In steady
state, substitution into the population dynamics leads to the bound
\(\mathcal{P}_{\text{out}} < \mathcal{P}_{\text{in}}\), even in the absence of additional loss
mechanisms. A detailed derivation of this constraint is provided in
appendix~\ref{app:pumprate}.

\subsection{Threshold: numerical results}

Typical parameters for simulating the Yb:YAG RBSL are shown in \autoref{TAB:parameters}.

\begin{figure}[t]
\includegraphics[width=0.7\columnwidth]{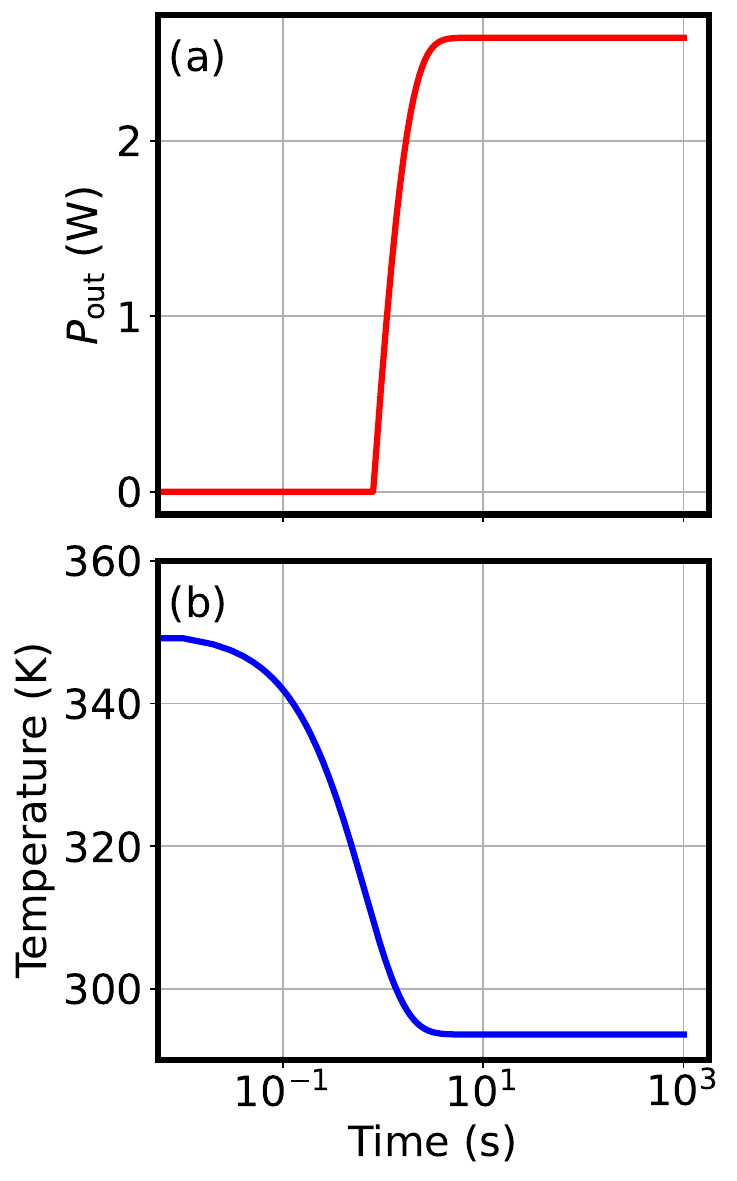}
\caption{
\textit{Delayed onset of lasing}. (a), (b) Plots for a representative case of RBSL operation with an initial temperature of $350\,\mathrm{K}$ and an input power of \(\mathcal{P}_{\text{in}}=25~\mathrm{W}\) to illustrate a delayed onset of lasing. The environment temperature is \(300\,\mathrm{K}\). Initially, the system lies above the lasing threshold temperature and therefore operates in a fluorescence-dominated cooling regime with no coherent output. As anti-Stokes cooling lowers the temperature from $350\,\mathrm{K}$, the system eventually crosses into the lasing region, at which point a finite laser output builds up. The emergence of lasing is therefore not immediate, but occurs only after the gain medium has self-cooled into the temperature window where the threshold condition is satisfied. All other model parameters are listed in \autoref{TAB:parameters}.
}
\label{Fig3}
\end{figure}

\begin{table}[b]
\centering
\caption{Model parameters used in the simulations.}
\begin{tabular}{lll}
\hline
Parameter & Value & Description \\
\hline

$r$ & $2~\mathrm{mm}$ & Crystal radius \\
$L$ & $1~\mathrm{mm}$ & Crystal thickness \\

$\rho$ & $4550~\mathrm{kg\,m^{-3}}$ & YAG mass density \\
$C_p$ & $590~\mathrm{J\,kg^{-1}K^{-1}}$ & YAG specific heat \\
$k$ & $10~\mathrm{W\,m^{-1}K^{-1}}$ & Thermal conductivity of YAG \\
$G_{\mathrm{th}}$ & $0.05~\mathrm{W\,K^{-1}}$ & Effective thermal conductance \\
$T_{\mathrm{env}}$ & $300~\mathrm{K}$ & Environment temperature \\

$n_{\mathrm{dop}}$ & $1.38\times10^{26}~\mathrm{m^{-3}}$ & Ion density \\

$g$ & $2\pi\times22.53~\mathrm{s^{-1}}$ & Light–matter coupling \\
$\gamma$ & $208.25~\mathrm{s^{-1}}$ & Spontaneous emission rate  \\
$\kappa$ & $10^7~\mathrm{s^{-1}}$ & Cavity decay rate \\
$\Gamma$ & $10^{10}~\mathrm{s^{-1}}$ & Thermalization rate \\

$\mathcal{P}_{\text{in}}$ & $25~\mathrm{W}$ & Incident pump power \\

\hline
\end{tabular}
\label{TAB:parameters}
\end{table}
A notable feature of a RBL that can be captured by the model is the delayed onset of lasing. A representative operating regime exhibiting this behavior is shown in \fref{Fig3}(a),(b), where we consider an input power of \(\mathcal{P}_{\text{in}}=25~\mathrm{W}\), an environment temperature of 300 K, and an initial gain-medium temperature of 350 K. For these parameters, the input power is insufficient to reach the lasing threshold at the initial temperature. However, since the gain medium operates in the solid-state cooling regime at this pump power, its temperature decreases over time until the lasing threshold is reached. At this point, a non-zero intracavity photon number builds up, leading to a finite laser output power.

\begin{figure*}[t]
\includegraphics[width=1.7\columnwidth]{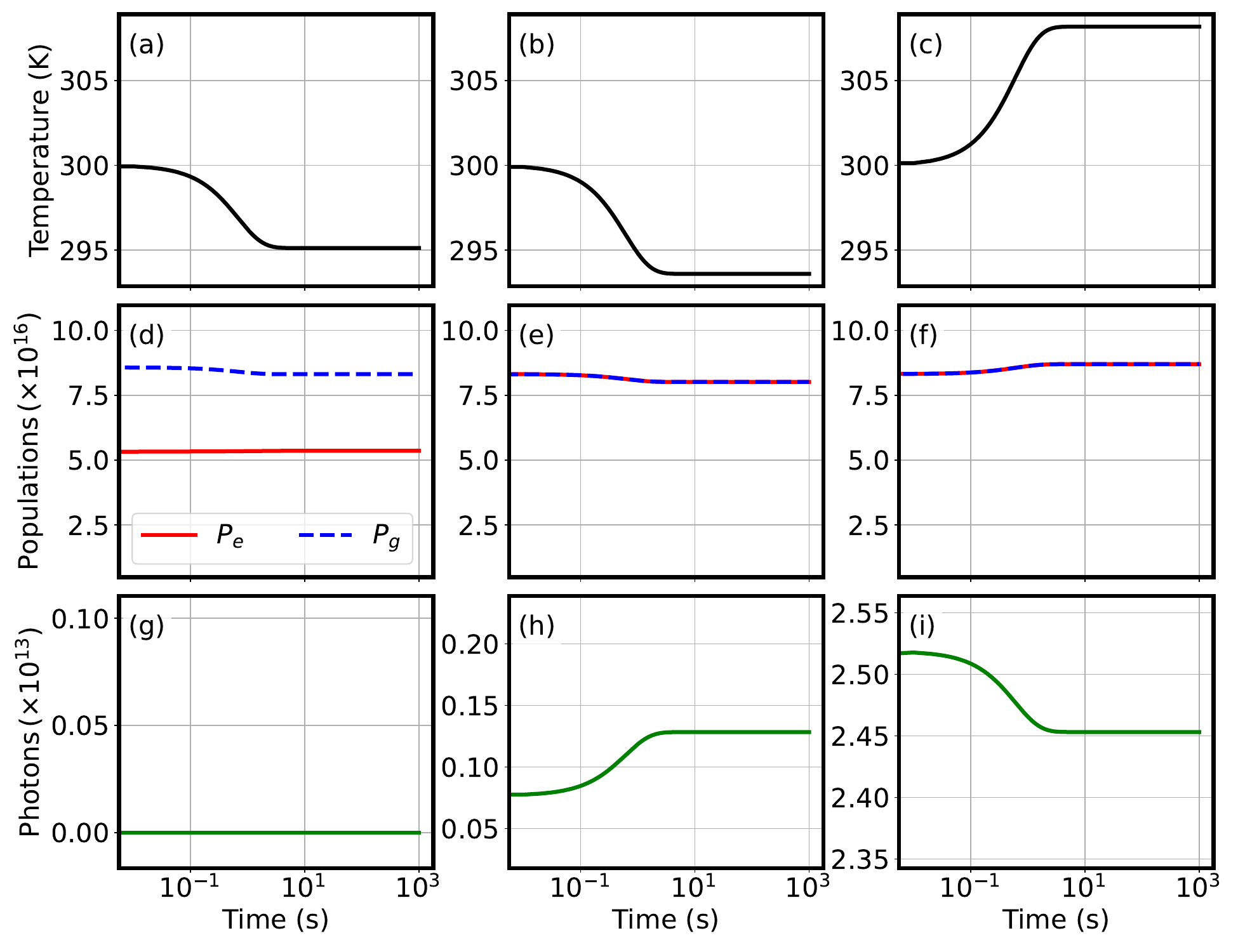}
\caption{
\textit{Regimes of RBSL operation.} Comparison of the two-level model for identical system parameters and varying input powers:
$\mathcal{P}_{\text{in}}=15$~W (a,d,g), $25$~W (b,e,h), and $75$~W (c,f,i).
The three columns correspond to distinct regimes: (i) cooling without population inversion or steady state lasing,
(ii) cooling with sufficient power to sustain lasing, and (iii) heating at high input power, where dissipative processes
overcome anti-Stokes cooling despite its dominance in the Yb:YAG pumping scheme.
The first row shows the temperature dynamics as the system approaches a new equilibrium, the second row shows
the level populations $P_e$ and $P_g$, and the third row shows the intracavity field amplitude $|\alpha|$.
In all cases, the initial and environmental temperatures are $300$~K and the thermal conductance is fixed at $G_{\text{th}}=0.05$.
}
\label{Fig4}
\end{figure*}

We also model three other common lasing regimes that can be practically achieved using Yb:YAG. In \fref{Fig4}, we model three distinct operating regimes, only changing the input power into the gain medium and otherwise maintaining the parameters given in \autoref{TAB:parameters}. The first one (left column, \(\mathcal{P}_{\text{in}}=15~\mathrm{W}\) input) corresponds to a configuration that never reaches the lasing regime but achieves solid-state cooling. The second one (middle column, \(\mathcal{P}_{\text{in}}=25~\mathrm{W}\) input) reaches a lasing regime while solid‑state cooling simultaneously occurs. The third one (right column, \(\mathcal{P}_{\text{in}}=75~\mathrm{W}\) input) reaches a lasing regime that stabilizes at a temperature above the environmental temperature, representing lasing with heat generation. In the left column, no population inversion is achieved so there are no photons present in the lasing cavity, and thus no lasing output. But because our level and pumping scheme is engineered to maximize anti-Stokes transitions, this input power still contributes to solid-state cooling which leaves the gain medium cooler than the environment. In the middle column, we have population inversion and also solid-state cooling. We note that population inversion is effectively clamped when temperature does not broadly vary, as we have seen in \eqref{eq:DeltaPth_main}. Finally, in the right column, we can see the effect of greatly increasing the input power and entering the regime of lasing while heating the gain medium relative to the surrounding environment. Entering this regime required a drastic increase in power due to only having one Stokes transition in the level scheme and due to the other heating mechanisms scaling with the number of photons and Beer-Lambert absorption in the YAG host.

\begin{figure}[t]
\includegraphics[width=0.9\columnwidth]{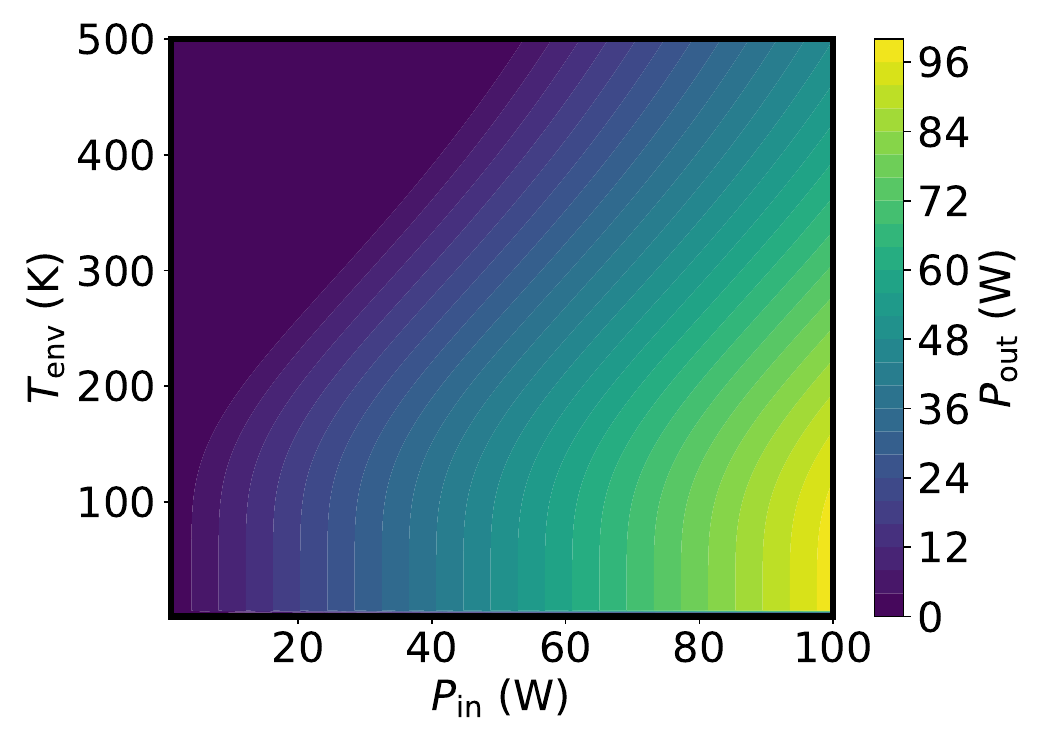}
\caption{
\textit{$\mathcal{P}_{\text{out}}$ vs $\mathcal{P}_{\text{in}}$ and $T_{\text{env}}$}. Contour plot of output power as the environment temperature and the input power are varied. The initial temperature is set equal to the environment temperature and all other model parameters are listed in \autoref{TAB:parameters}.
}
\label{Fig5}
\end{figure}

The relationship between input power and output lasing power is of main interest to gauge the potential of a solar laser. With the environment temperature, $T_{\text{env}}$, as an additional control, we can simulate a variety of operating conditions for a solar laser, shown in the contour plot in \fref{Fig5}. The consistent parameters used across all simulations in this plot are those shown in \autoref{TAB:parameters}. At lower temperatures, lasing output power is high, asymptotically approaching the input pump power, and fluorescence power is low. At higher temperatures, the laser output power is low and fluorescence power is high. In the contour plot, the initial temperature used for each simulation is equal to the environment temperature, and the lasing output power is calculated from the steady state intracavity photon number. 

\section{Discussion}
A central prediction of the present work is the emergence of three distinct dynamical regimes within a single physical device architecture. Depending on the interplay between optical driving, cavity feedback, and thermal processes, the system can operate in (i) a purely cooling regime without lasing, where no intracavity field builds up; (ii) a radiation-balanced lasing regime, in which coherent emission coexists with net cooling; and (iii) a heating-dominated lasing regime, where optical absorption and non-radiative processes overcome anti-Stokes extraction.

What distinguishes the present framework is that the transitions between these regimes are not governed solely by a global heat balance, but crucially depend on temperature-controlled population redistribution among internal sublevels. This introduces an additional layer of structure: the lasing threshold is no longer a fixed function of pump power alone, but acquires an intrinsic temperature dependence through the statistical occupation of manifolds. As a result, mapping the boundary between lasing and non-lasing operation as a function of environmental temperature is predicted to reveal a nontrivial, curved phase boundary, reflecting the underlying redistribution dynamics.

A second key signature is the possibility of delayed lasing activation driven by self-cooling. If the system is initially prepared at a temperature above the effective lasing threshold, anti-Stokes fluorescence can reduce the internal temperature until the threshold condition is met. At that point, the intracavity field rapidly builds up and coherent emission sets in. This leads to a characteristic temporal evolution: an initial regime dominated by broadband fluorescence and cooling, followed by a sharp transition to strong, coherent laser output once the population distribution crosses the threshold condition.

A third experimentally accessible signature is a correlated trade-off between fluorescence and laser emission near thermal equilibrium. As the system approaches the environmental temperature, anti-Stokes fluorescence becomes the dominant dissipation channel, while the coherent laser output correspondingly diminishes. This implies a measurable anticorrelation between broadband fluorescence intensity and narrowband lasing power under controlled parameter sweeps. Such behavior connects naturally to existing experimental studies of radiation-balanced lasing in solid-state systems, including Yb:YAG-based platforms, where precise thermal management and profiling at the balance point play a central role.

\section{Conclusions}

We have developed a microscopic open-quantum-system theory of a RBSL, in which optical gain and thermal dynamics are treated on equal footing. Starting from a Lindblad description of a multi-level Yb$^{3+}$ ion coupled to a cavity mode, we constructed a closed dynamical framework that links population dynamics, optical coherence, and lattice temperature through explicit energy-exchange channels. In this formulation, temperature is not an externally prescribed parameter but a dynamical variable that feeds back onto the lasing process through Boltzmann redistribution among electronic sublevels.

A central result of this work is the identification of a microscopic \emph{temperature--gain feedback mechanism}. Thermal redistribution controls the fraction of ions participating in the lasing transition and thus directly modifies the available inversion and threshold condition, while the optical dynamics simultaneously reshape the temperature through a competition between anti-Stokes fluorescence cooling and several heating channels. This closed-loop coupling gives rise to behavior that is absent in conventional rate-equation or purely thermodynamic descriptions of RBLs.

By exploiting the separation of timescales between fast intra-manifold thermalization and the slower optical processes of cavity decay, spontaneous emission, and the incoherent pump, we derived a reduced temperature-dependent Maxwell--Bloch model that retains the essential microscopic structure through partition functions and detailed-balance relations. This reduction yields analytical threshold conditions and makes explicit how temperature reshapes the gain landscape. Within this framework, we predict distinct operating regimes, including cooling without lasing, lasing with net cooling, and lasing with net heating, as well as transient effects such as delayed lasing onset induced by self-cooling into threshold.

Our analysis also shows that radiation-balanced lasing imposes stringent microscopic constraints on the gain medium. In particular, sufficiently small spontaneous-emission rates are required to reach threshold at realistic pump powers, making rare-earth systems with long-lived excited states not merely advantageous but effectively necessary for practical implementations. This identifies a material-level criterion for RBSL operation that complements earlier thermodynamic and system-level considerations.

The minimal model developed here is intended to highlight the microscopic mechanisms underlying a RBSL, while remaining simple enough to yield analytical expressions for key quantities such as the lasing threshold and steady state output power. For device-specific quantitative predictions, this framework can be systematically refined by replacing the idealized equal-spacing and equal-branching assumptions with spectroscopically resolved Stark splittings, transition strengths, and absorption and emission cross sections for the chosen host crystal. Furthermore, to accurately describe laser configurations with gain media of different sizes in a non‑ideal optical cavity, the model must incorporate additional variables such as optical reabsorption, radiation trapping, and pump depletion, all of which can influence the effective cooling power and the lasing threshold.

It will also be important to move beyond a lumped-temperature description and include spatial heat transport, thermal gradients, and possible thermal-lensing effects, especially in high-power or large-aperture solar geometries. Interestingly, our approach paves the way towards a quantum-optical treatment of structured photonic environments and selective Purcell cavity engineering to enhance fluorescence without proportionally suppressing laser performance.

These extensions would turn the present framework from a minimal microscopic theory into a quantitatively predictive design tool for realistic RBSLs. Even in its current form, however, the theory clearly establishes the key physical principle: RBSL operation is governed by a nontrivial interplay between microscopic level structure, thermal redistribution, and cavity-mediated gain. By placing this interplay on a microscopic open-system footing, the present work underscores both the opportunities and the limitations of thermally self-stabilized solar lasers and provides a foundation for their systematic optimization.\\

\section*{Author contributions}
M.K and H.F conceived the original idea of the RBSL and carried out the initial Yb:YAG study. The overall theoretical modeling strategy was defined jointly by H.F, M.K, and C.G. The microscopic theoretical framework based on open quantum system dynamics, together with the analytical derivations and calculations, was developed by A.J and C.G. The manuscript was written by A.J under the supervision of J-M.M and C.G, with feedback from H.F and M.K.
\section*{Acknowledgments}
We acknowledge financial support from the Max Planck Society. This work was supported by the Natural Sciences and Engineering Research Council of Canada (RGPIN-2023-05365). 
\bibliography{Bibliography}

\appendix

\onecolumngrid

\newpage

\section{Reduction from the seven-level model to a two-manifold model} \label{app:7to2}

We assume strong intra-manifold thermalization (typically on timescales of the order of picoseconds),
\(
\Gamma \gg \{\gamma,\kappa,R\},
\)
so that the populations within each manifold rapidly reach a Boltzmann thermal equilibrium. Population ratios within each manifold can be defined with Boltzmann factors, \(q=\bar n/(\bar n+1)\). For each manifold we can define total populations, \(N_e\) and \(N_g\) and partition functions \(Z_e\) and \(Z_g\) as follows: 
\begin{itemize}
    \item \noindent \textbf{Ground electronic manifold:}
For the chain \(g_0 \leftrightarrow g_1 \leftrightarrow i \leftrightarrow g\),
\[
P_{g_1}=qP_{g_0},\qquad P_i=q^2P_{g_0},\qquad P_g=q^3P_{g_0}.
\]
Defining the total ground-manifold population \(N_g\),
\[
N_g = P_{g_0}+P_{g_1}+P_i+P_g,
\]
yields \(N_g=P_{g_0}Z_g\) with
\begin{equation}
Z_g = 1+q+q^2+q^3,
\label{eq:SI_Zg}
\end{equation}
and therefore
\begin{equation}
P_i=\frac{q^2}{Z_g}N_g,\qquad
P_g=\frac{q^3}{Z_g}N_g.
\label{eq:SI_PiPg_from_Ng}
\end{equation}

\item \noindent \textbf{Excited electronic manifold:}
For the chain \(e \leftrightarrow e_1 \leftrightarrow e_2\),
\[
P_{e_1}=qP_e,\qquad P_{e_2}=q^2P_e.
\]
Defining the total excited-manifold population
\[
N_e = P_e+P_{e_1}+P_{e_2},
\label{eq:SI_Ne_def}
\]
gives \(N_e=P_eZ_e\) with
\begin{equation}
Z_e = 1+q+q^2,
\label{eq:SI_Ze}
\end{equation}
and therefore
\begin{equation}
P_e=\frac{N_e}{Z_e}.
\label{eq:SI_Pe_from_Ne}
\end{equation}
\end{itemize}
Conservation of populations involves
\begin{equation}
N_g+N_e=N \quad\Rightarrow\quad N_g=N-N_e.
\label{eq:SI_conservation}
\end{equation}
The full equations of motion for the 7-level model coupled to the cavity field are listed below.
\begin{subequations}
\begin{align}
\dot{\alpha}
&= -\frac{\kappa}{2}\,\alpha + g\,S , \\
\dot{S}
&= g\,\alpha\,(P_e - P_g)
- \left[ \tfrac{1}{2}\Gamma(2\bar n + 1) + 2\gamma \right] S , \\
\dot{P}_{g_0}
&= \gamma (P_e + P_{e_1} + P_{e_2})
- \Gamma \bar n\, P_{g_0}
+ \Gamma (\bar n + 1)\, P_{g_1}, \\
\dot{P}_{g_1}
&= \gamma (P_e + P_{e_1} + P_{e_2})
+ \Gamma \bar n\, (P_{g_0}-P_{g_1}) 
+ \Gamma (\bar n + 1)\, (P_i-P_{g_1}) , \\
\dot{P}_i
&= \gamma (P_e + P_{e_1} + P_{e_2})
- R\,P_i
+ \Gamma \bar n\, (P_{g_1}-P_i)
+ \Gamma (\bar n + 1)\, (P_g-P_i) , \\
\dot{P}_g
&= \gamma (P_e + P_{e_1} + P_{e_2})
+ g(\alpha^\ast S + \alpha S^\ast)
+ \Gamma \bar n\, P_i 
- \Gamma (\bar n + 1)\, P_g, \\
\dot{P}_e
&= R\,P_i
- g(\alpha^\ast S + \alpha S^\ast)
- (4\gamma + \Gamma \bar n)\, P_e
+ \Gamma (\bar n + 1)\, P_{e_1}, \\
\dot{P}_{e_1}
&= - 4\gamma\, P_{e_1}
+ \Gamma \bar n\, (P_e - P_{e_1})
+ \Gamma (\bar n + 1)\, (P_{e_2} - P_{e_1}) , \\
\dot{P}_{e_2}
&= - 4\gamma\, P_{e_2}
+ \Gamma \bar n\, P_{e_1}
- \Gamma (\bar n + 1)\, P_{e_2}.
\end{align}
\end{subequations}

The equations above are cumbersome and do not offer much insight into the competition of processes leading to lasing and cooling. Instead, we will proceed with simplifying our model by a series of approximations. The full seven-level model is kept as a reference to benchmark the correctness of the employed approximations. 

From the 7-level equations, we know that the important contributions to the incoherently pumped solar laser come from the Bloch equations for the cavity mode, \(\alpha\), the coherence for the lasing transition \(S\), and the populations. Due to conservation of total population, we need only to consider populations within a single manifold because we can infer the population relationship of the other manifold; meaning that \(\dot{N_e}=-\dot{N_g}\). Furthermore, considering an entire manifold as the Bloch equation for population simplifies the relevant interactions because thermalization within a manifold is balanced and the thermalization terms will cancel out. So given the definition of \(N_e\), we derive
\begin{align*}
\dot{N}_e= \dot{P}_e + \dot{P}_{e_1} + \dot{P}_{e_2} \;\Rightarrow\; \dot{N}_e= R\,P_i - 4\gamma\, N_e - g(\alpha^\ast S + \alpha S^\ast).
\end{align*}
Finally, with \(P_i\) and \(P_g\) reconstructed using \eqref{eq:SI_PiPg_from_Ng} and \eqref{eq:SI_conservation}, and \(P_e\) via \eqref{eq:SI_Pe_from_Ne}, the system of equations reduces to
\begin{subequations}\label{eq:SI_reduced_system}
\begin{align}
\dot\alpha
&= -\frac{\kappa}{2}\alpha + gS,\\[4pt]
\dot S
&= -\Gamma_S S
+ g\alpha\left(\frac{N_e}{Z_e}-\frac{q^3}{Z_g}(N-N_e)\right),\\[6pt]
\dot N_e
&= R\frac{q^2}{Z_g}(N-N_e) - 4\gamma N_e
- g(\alpha^\ast S+\alpha S^\ast).
\end{align}
\end{subequations}
This is a standard microscopic model for a lasing system in which the cavity field is driven by the collective coherence $S$, while $S$ itself is generated by the population inversion $N_e$. The description is at the mean-field level and allows one to identify a lasing threshold via a linear stability analysis around the trivial (non-lasing) solution, i.e., by introducing an infinitesimal seed. \textit{The threshold is reached when an arbitrarily small seed in the cavity field amplitude is exponentially amplified and evolves toward a finite steady state value that is independent of the initial seed magnitude.} Temperature enters implicitly through the system variables and explicitly through the effective damping rate
\[
\Gamma_S = \Gamma(\bar n + 1/2) + 2\gamma.
\]

\begin{figure*}[b]
\includegraphics[width=0.6\columnwidth]{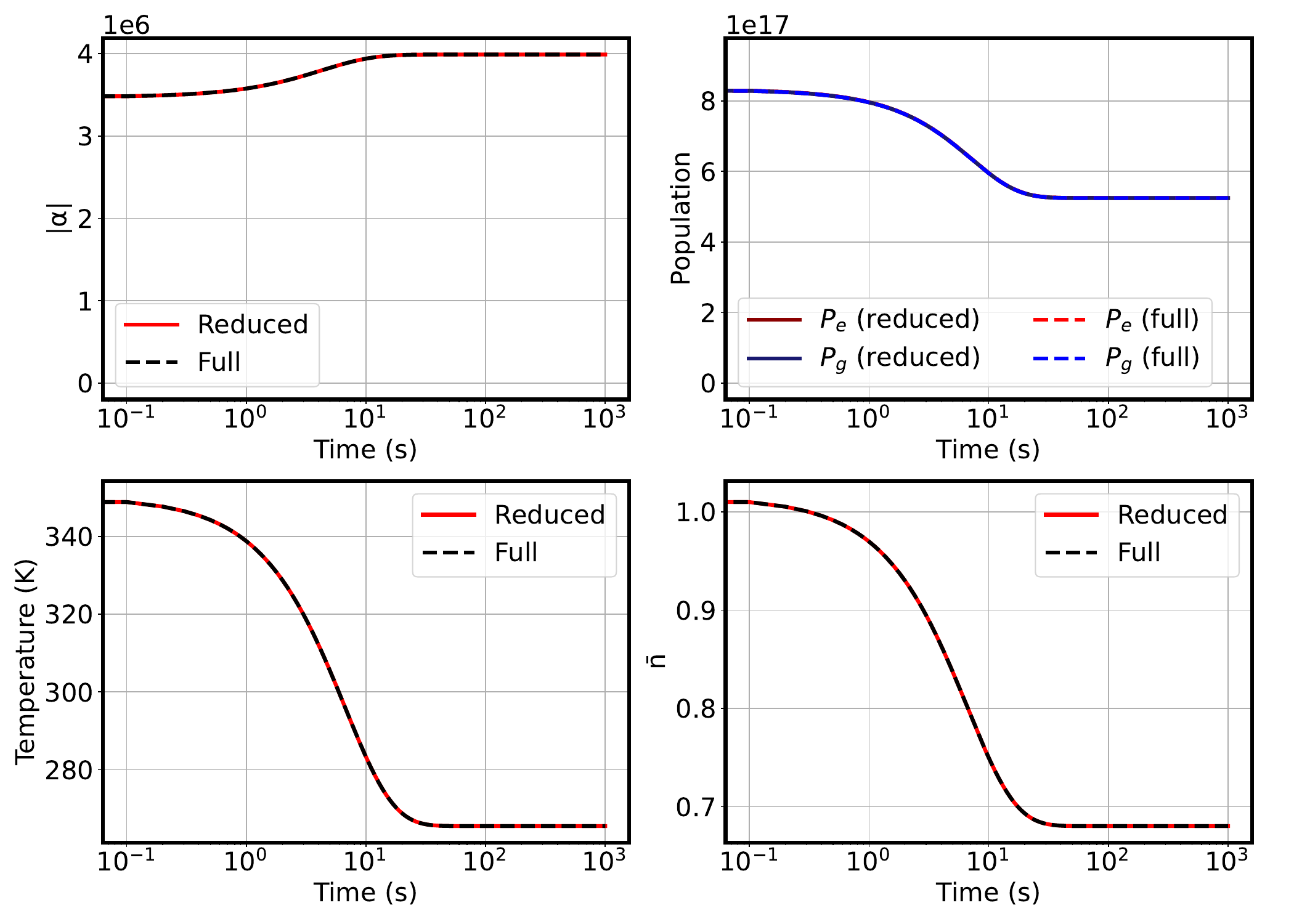}
\caption{
Comparison of various dynamics of the seven-level (full, exact) model and two-level (reduced) model with an incoherent drive. Agreement is excellent owing to the fact that $\Gamma$ dominates all other rates in the system. The values of the rates are $\Gamma=1\times10^{10}$,  $\kappa=1\times10^{9}$, $\gamma=5\times10^{3}$, and $R=2.5\times10^{4}$.
}
\label{FIG10}
\end{figure*}

\section{Analytical derivation of the lasing threshold in the simplified two-level model} \label{app:thresholds}

Let us now derive an analytical scaling for the lasing threshold. While numerically it is straightforward to check for the amplification of a small seed, analytically we will simply ask for the conditions under which the steady states shows \(|\alpha|^2>0\). We assume fixed \(\bar n\) (hence fixed \(q,Z_g,Z_e,\Gamma_S\)) and use the steady state version of \eqref{eq:SI_reduced_system}. The upshot is the threshold number of total ions \(N_\text{th}\) required for lasing initialization, from which we can directly obtain the threshold pumping rate \(R_\text{th}\). We begin by setting \eqref{eq:SI_reduced_system} to steady state as
\begin{subequations}\label{eq:SteadyStateSystem}
\begin{align}
0
&= -\frac{\kappa}{2}\,\alpha + g\,S,
\label{eq:SteadyStateOne}\\[4pt]
0
&= -\Gamma_S\, S
+ g\,\alpha\left(
\frac{N_e}{Z_e}
- \frac{q^3}{Z_g}(N - N_e)
\right),
\label{eq:SteadyStateTwo}\\[6pt]
0
&= R\frac{q^2}{Z_g}(N - N_e)
- 4\gamma\, N_e
- g(\alpha^\ast S + \alpha S^\ast).
\label{eq:SteadyStateThree}
\end{align}
\end{subequations}
From \eqref{eq:SteadyStateOne} we can obtain the threshold value of \(S\) to be
\begin{equation}
    S=\frac{\kappa}{2g}\alpha.
\end{equation}
Replacing this into \eqref{eq:SteadyStateThree} leads to the following expression for \(|\alpha|^2>0\) 
\begin{align}
\kappa|\alpha|^2 =& R\frac{q^2}{Z_g}(N-N_e) - 4\gamma N_e.
\end{align}
Requiring that \(|\alpha|^2>0\), leads immediately to
\begin{equation}
R\frac{q^2}{Z_g}(N - N_e) > 4\gamma N_e
\;\Rightarrow\;
N > \left(\frac{4\gamma Z_g}{R q^2} + 1\right) N_e.
\label{eq:ConditionNwithNe}
\end{equation}
Now we only require the steady state value of \(N_e\), which we can obtain from manipulating \eqref{eq:SteadyStateTwo} by replacing $S=\kappa/(2g)\alpha$ and requiring again that $|{\alpha}|>0$ to get
\begin{align}
N_e
&= \left(
\frac{\kappa \Gamma_S}{2g^2}
+ \frac{q^3}{Z_g}\, N
\right)
\left(
\frac{1}{Z_e}
+ \frac{q^3}{Z_g}
\right)^{-1}.
\label{eq:SSNe}
\end{align}
Finally, we can obtain the threshold condition by substituting \eqref{eq:SSNe} into \eqref{eq:ConditionNwithNe} and solving for values of \(N\) that satisfy the inequality. This leads to a threshold value for the incoming pump rate
\begin{equation}
R_{\mathrm{th}}
=
\frac{4\gamma Z_e\!\left(\dfrac{\kappa \Gamma_S}{2g^2}\, Z_g + N q^3\right)}
{q^2\!\left(N - \dfrac{\kappa \Gamma_S}{2g^2}\, Z_e\right)}.
\label{eq:SI_Rth}
\end{equation}
Note that the factor \(\kappa\Gamma_S / 2g^2\) comes from solving \eqref{eq:SteadyStateTwo} and is equal to the population difference of \(P_e-P_g\) in steady state. Let us also remark on this analytical expression of the threshold pump rate: the expression is only valid as long as $N>\kappa \Gamma_S/(2g^2\, Z_e)$ keeping the denominator positive and finite. The requirement comes from our requirement that there is a positive cavity field amplitude, $|\alpha|>0$, which is only true for above threshold, i.e. when the number of gain ions is large enough. The expression is an approximation that fits very well, well above the threshold $N$. Also, notice that the threshold number of ions depends weakly on temperature via $Z_e$ which goes from $1$ at low temperature to $3$ at very high $T$.

\section{Connecting the incoherent pumping rate ($R$) to the incident solar radiation intensity}
\label{app:pumprate}
The equations of motion simulate the injection of energy into the laser gain medium as a number of excitations per second; in \eqref{eq:SI_reduced_system}, this injection term is \(RP_i=R\frac{q^2}{Z_g}(N-N_e)\). It is important to constrain the value of \(R\) to be directly calculated from the incident solar radiation \(\mathcal{P}_{\text{in}}\) for several reasons. One reason is to prevent using pump rates that would not be achievable from the incident solar radiation. Another reason is to constrain the output laser power to not exceed the input power. Finally, the pump rate in the gain medium will vary while the system reaches equilibrium rather than remain constant. While there are many conditions and factors which limit the conversion efficiency of the incident solar radiation to the absorbed power of the gain medium (Beer-Lambert, multipass in the cavity, absorption spectra of the Yb ion, etc) we can first directly consider the extreme case of the absorbed power being equal to \(\mathcal{P}_{\text{in}}\). In this case we can define the rate \(R\) [1/s] from the power [J/s] divided by the total energy of the photons available to be pumped [J] from population \(P_i\). The rate can then be expressed as

\begin{equation}
    R(t) = \frac{\mathcal{P}_{\text{in}}}{\hbar\omega_p P_i(t)},
    \label{eq:DynamicRfromPin}
\end{equation}

where we note that the population dependence on time makes the pump rate dynamic. 

Let us check for consistency, that the output power is bounded by the incoming power, in the absence of other sources of loss. By substituting \eqref{eq:DynamicRfromPin} into the equations of motion we can directly apply the constraint which prevents the output power from exceeding the input power. This can be seen from the steady state equation for the evolution of the excited state populations, 
\begin{equation*}
    \dot{N_e} = 0 = RP_i - 4\gamma N_e - g(\alpha^\ast S+\alpha S^\ast).
\end{equation*}
Knowing that in steady state, \(S = (\kappa/2g)\alpha\) and the cavity output is \(\mathcal{P}_{\text{out}}=\kappa\hbar \omega_c|\alpha|^2\), we substitute to obtain

\begin{align*}
    \mathcal{P}_{\text{out}} = & \frac{\omega_c}{\omega_p}\mathcal{P}_{\text{in}} - 4\gamma N_e \hbar \omega_c.
\end{align*}
This shows that power is lost into fluorescence but also that the output power is bounded by the ratio of \(\omega_c/\omega_p<1\) which enforces \(\mathcal{P}_{\text{out}}<\mathcal{P}_{\text{in}}\). 

\section{Other sources of heating: thermal coupling to the external environment}
\label{app:enviroheat}

In addition to the internal optical heating and cooling mechanisms
discussed in the main text, the gain medium exchanges heat with its
macroscopic environment. For a solid crystal, the three relevant
mechanisms are: (i) conduction to a mount or heat sink, (ii) convection
to surrounding gas (if present), and (iii) thermal radiation. These are the dominant independent macroscopic heat-transfer channels.

For a cylindrical disk of radius $r$ and thickness $L$, the heat
capacity is
\begin{equation}
C_{\mathrm{th}} = \rho C_p V
= \rho C_p \pi r^2 L ,
\end{equation}
where $\rho$ and $C_p$ are the mass density and specific heat of YAG.

The environmental heat flow can be written as
\begin{equation}
\mathcal{P}_{\mathrm{env}}(T)
=
\mathcal{P}_{\mathrm{cond}}
+
\mathcal{P}_{\mathrm{conv}}
+
\mathcal{P}_{\mathrm{rad}} .
\end{equation}

\paragraph{Conduction.}
If the disk is bonded to a heat sink across one face,
Fourier’s law gives
\begin{equation}
\mathcal{P}_{\mathrm{cond}}
= G (T - T_{\mathrm{env}}),
\qquad
G \simeq \frac{k A}{L},
\end{equation}
where $k$ is the thermal conductivity of YAG and
$A=\pi r^2$ is the contact area.
For YAG ($k \sim 10~\mathrm{W\,m^{-1}K^{-1}}$)
and typical dimensions $r=2~\mathrm{mm}$, $L=1~\mathrm{mm}$,
we obtain
\[
G \sim 0.1~\mathrm{W/K}.
\]

\paragraph{Convection.}
If the crystal is surrounded by air,
\begin{equation}
\mathcal{P}_{\mathrm{conv}}
= h A_s (T - T_{\mathrm{env}}),
\end{equation}
where $h \sim 5$–$20~\mathrm{W\,m^{-2}K^{-1}}$
for natural convection and
$A_s$ is the surface area.
For the same disk geometry, $h A_s \sim 10^{-3}~\mathrm{W/K}$,
which is typically two orders of magnitude smaller than the
conductive coupling to a mount.

\paragraph{Thermal radiation.}
Thermal emission follows the Stefan–Boltzmann law,
\begin{equation}
\mathcal{P}_{\mathrm{rad}}
=
\epsilon \sigma A_s (T^4 - T_{\mathrm{env}}^4),
\end{equation}
which linearizes near $T_{\mathrm{env}}$ as
\begin{equation}
\mathcal{P}_{\mathrm{rad}}
\approx
4 \epsilon \sigma A_s T_{\mathrm{env}}^3
(T - T_{\mathrm{env}}).
\end{equation}
At room temperature this corresponds to an effective
conductance of order $10^{-3}~\mathrm{W/K}$ for millimeter-scale
crystals, again much smaller than conductive mounting.

\medskip

Therefore, for thin-disk geometries mechanically coupled to a heat
sink, conduction dominates the environmental heat exchange,
and the macroscopic thermal coupling can be well approximated as
\begin{equation}
\mathcal{P}_{\mathrm{env}}(T)
\simeq
G (T - T_{\mathrm{env}}).
\end{equation}
Only in vacuum operation without mechanical contact does thermal
radiation become the leading external heat-transfer mechanism.

\begin{figure*}[h]
\includegraphics[width=0.5\columnwidth]{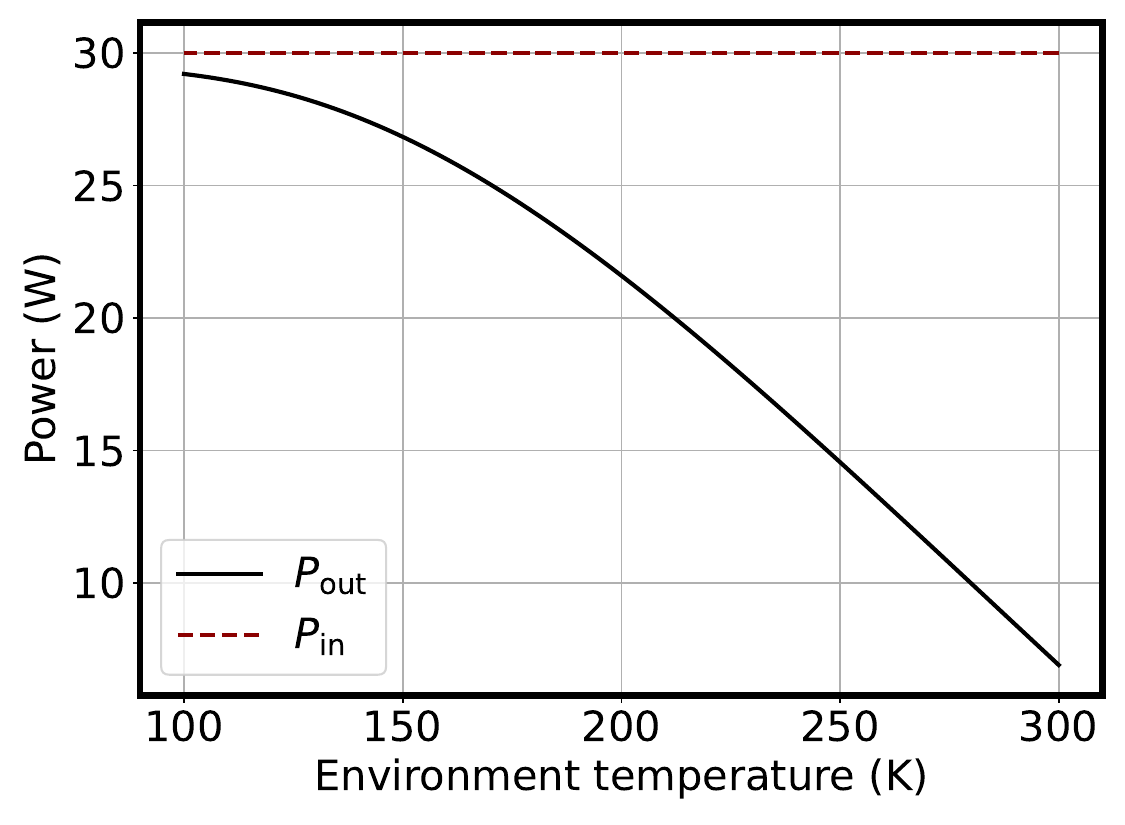}
\caption{
Power versus environment temperature.
}
\label{FIG12}
\end{figure*}

\end{document}